\documentclass[a4paper,11pt]{article}
\usepackage{pos}

\newcommand{\tB}{{\rm B}}
\newcommand{\tQ}{{\rm Q}}
\newcommand{\tS}{{\rm S}}

\newcommand{\hmuQ}{{\hat{\mu}_{\rm Q}}}

\newcommand{\hmuB}{{\hat{\mu}_{\rm B}}}

\title{QCD Equation of State with Strong Magnetic Fields and Nonzero Baryon Density}
\ShortTitle{QCD EoS with Strong Magnetic Fields and Nonzero Baryon Density}

\author[1]{Heng-Tong Ding}
\author[1]{Jin-Biao Gu}
\author*[1]{Arpith Kumar}
\author[1]{Sheng-Tai Li}

\affiliation[1]{Key Laboratory of Quark and Lepton Physics (MOE) and Institute of
    Particle Physics, \\
    Central China Normal University, Wuhan 430079, China}    

\emailAdd{arpithk@ccnu.edu.cn}

\abstract{In this work, we have carried out lattice simulations of $(2+1)$-flavor QCD using highly improved staggered quarks at the physical pion mass on $32^3 \times 8$ and $48^3 \times 12$ lattices, with magnetic field strengths ranging up to 0.8 GeV$^2$ and nonzero baryon chemical potentials employing the Taylor expansion framework. We present lattice QCD continuum estimate results, along with the magnetized hadron resonance and ideal gas comparisons, for the leading-order Taylor expansion coefficients for bulk thermodynamic quantities such as pressure, number density, energy density, and entropy density, focusing on the significant impact of strong magnetic fields.}

\FullConference{The 41st International Symposium on Lattice Field Theory (LATTICE2024)\\
 28 July - 3 August 2024\\
Liverpool, UK\\}

\begin{document}
\maketitle

\section{Introduction \label{sec:intro}}

The equilibrium properties of a thermodynamic system are fundamentally characterized by its equation of state (EoS), which encodes the relationship between key physical observables like pressure, energy density, and entropy density. Of particular interest is the behaviour of these energy-related observables under varying control parameters of the system, such as temperature and chemical potential, as well as external influences such as magnetic fields. Interestingly, in various settings at different physical scales, magnetic fields are expected to modify the EoS and its physical implications. For instance, the cosmological models suggest cosmic magnetic fields from density perturbations impacting the Friedmann equations \cite{Vachaspati:1991nm}. Another setting is magnetars, with the strongest natural magnetic fields influencing Tolman-Oppenheimer-Volkoff equations and thereby mass, radius, and energy density relations \cite{Duncan:1992hi, Harding:2006qn}. One of the most intriguing settings is off-central heavy-ion collisions, where high-velocity spectator-charged particles generate strong magnetic fields with strength $eB$ comparable to the strong interaction $\Lambda_{\rm QCD}^2$ scale \cite{Deng:2012pc, Skokov:2009qp}. In the presence of such strong magnetic fields, the underlying nontrivial topology of QCD can manifest multiple macroscopic phenomena, the most famous being the chiral magnetic effect \cite{Kharzeev:2007jp,Kharzeev:2020jxw}. The quest for such effects has triggered intensive ongoing experimental and theoretical studies \cite{Fukushima:2009ft,Fu:2013ica,STAR:2021mii,Kharzeev:2022hqz} (for a recent review, see \cite{Endrodi:2024cqn}). 

Lattice-regularized QCD using Taylor expansion for nonzero density has recently gained further interest due to its applicability in incorporating magnetic fields, where no sign problem arises. These studies suggest that strong magnetic fields can significantly influence QCD properties; in particular, the QCD thermodynamics \cite{Bali:2014kia}, its phase diagram \cite{Bali:2011qj,Ding:2020inp}, transport properties \cite{Astrakhantsev:2019zkr}, in-medium hadron properties \cite{Bonati:2015dka, Endrodi:2019whh} etc. However, still, much less is known about the details of the changes in the degrees of freedom in QCD with external magnetic fields. While phenomena such as magnetic catalysis, its inverse, and the reduction of the pseudo-critical temperature ($T_{pc}$) under magnetic fields have been extensively studied in the context of chiral condensates~\cite{DElia:2011koc,Bali:2012zg,Ding:2020hxw,Ding:2025pbu,Ding:2022tqn}, the chiral condensate itself remains a theoretical quantity that cannot be directly measured in experiments. Both theoretically and experimentally accessible, fluctuations of and correlations among net baryon number (B), electric charge (Q) and strangeness (S) have been extensively employed to probe changes of degrees of freedom in zero magnetic fields and thereby construct QCD EoS \cite{HotQCD:2012fhj,Bazavov:2020bjn,Bollweg:2021vqf,Bazavov:2017dus,Bollweg:2022fqq}. Recently, such lattice studies in the presence of magnetic fields are gradually unfolding  \cite{Ding:2021cwv,Ding:2023bft,Borsanyi:2023yap,Astrakhantsev:2024mat, Borsanyi:2025mrf}. This work aims to further bridge the gap in understanding the magnetic influence on the QCD EoS. Focusing on the heavy ion collision scenario, incorporating strangeness neutrality and isospin asymmetry, we construct the EoS with appropriate combinations of susceptibilities in the physical conserved charge basis.

This proceeding is organized as follows. In Sec. \ref{sec:CCTE}, we briefly introduce the lattice QCD setup relevant to our simulations, the extent of physical parameters, and the basic framework of Taylor series expansion. Before going on to the equation of state, Sec. \ref{sec:initial_cond} discusses strangeness neutrality and isospin asymmetry constraints with a fixed net electric charge to baryon number ratio relevant to initial and freeze-out conditions in heavy-ion collisions. Furthermore, we sketch physical interpretations of our lattice results with the help of the hadron resonance gas (HRG) and ideal gas model in the context of magnetic fields. Finally, in Sec. \ref{sec:results}, we present our lattice results for QCD magnetic EoS.

\section{Lattice setup and thermodynamics in magnetic fields}
\label{sec:CCTE}

We consider (2+1)-flavor QCD with highly improved staggered quarks (HISQ) \cite{Follana:2006rc} and a tree-level improved Symanzik gauge action extensively utilized by the HotQCD Collaboration \cite{Bazavov:2011nk,HotQCD:2014kol,Bazavov:2017dus,HotQCD:2018pds,Bazavov:2019www,HotQCD:2019xnw}. For thermodynamical systems, one of the most fundamental quantities is pressure, which in the grand canonical ensemble can be expressed in terms of the logarithm of partition function as $ {p} = (T/V)\ln Z (eB, T, V, {\mu})$. However, lattice study of strongly interacting matter at nonzero density encounters the infamous sign problem, which can be circumvented by Taylor expansion of the pressure, for instance, in the physical conserved charge basis,
\begin{equation}
    \hat{p} \equiv \frac{p}{T^4} = \sum_{ijk}\frac{1}{i!j!k!}~ \chi^{\tB \tQ \tS}_{ijk}~ \hat{\mu}^{i}_{\tB} \hat{\mu}^{j}_{\tQ} \hat{\mu}^{k}_{\tS}; \qquad
    \chi^{\rm BQS}_{ijk} =\frac{1}{VT^3} \left( \frac{\partial}{\partial  \hat{\mu}_{\rm B}} \right)^i \left( \frac{\partial}{\partial  \hat{\mu}_{\rm Q}} \right)^j \left( \frac{\partial}{\partial  \hat{\mu}_{\rm S}} \right)^k \ln Z\Bigg|_{\hat{\mu}_{\rm B,Q,S}=0}\,,
    \label{eq:suscp_uds}
\end{equation}
where $\chi^{\rm BQS}_{ijk}$ represents the generalized susceptibilities and the leading-order $i+j+k=2$ corresponds to the fluctuations of and correlations among them. In principle, we first compute susceptibilities in the fundamental quark chemical potentials basis \cite{Allton:2002zi,Gavai:2003mf}, and then can go back and forth with the physical basis using the following relations
\begin{align}
    \mu_u = \frac{1}{3} \mu_\tB + \frac{2}{3} \mu_\tQ,~~ 
\mu_d = \frac{1}{3} \mu_\tB - \frac{1}{3} \mu_\tQ, ~~
\mu_s = \frac{1}{3} \mu_\tB - \frac{1}{3} \mu_\tQ - \mu_\tS \quad \to \quad    
\chi^{uds}_{ijk} \leftrightarrow \chi^{\text{BQS}}_{ijk}.
\end{align}
We introduce a magnetic field on lattice, defined as curl of vector potential, $\Vec{B} = \Vec{\nabla}\times \Vec{A}$, implying no associated sign problem.  We consider a constant magnetic field along the $z$ direction and fix the gauge using the Landau gauge, leading to fixed factors $u_{\mu}(n)$ of the U(1) field \cite{Ding:2021cwv, Ding:2020hxw}. From Stokes theorem, the quantized nature of magnetic field on a finite lattice (the area is also quantized) leads to corresponding strength $eB$ quantization
\begin{equation}
eB = {6\pi N_b}/a^{2}{N_x N_y}, \quad 0\leq N_b \in \mathbb{Z} < {N_x N_y}/{4}\,,
\end{equation}
where $N_b$ denotes the number of magnetic fluxes through $x-y$ plane unit area and is further constrained by the periodic boundary condition for U(1) links for all except the $x$ direction. 

Our simulations are performed on spatially symmetric lattices with fixed spatial-to-temporal aspect ratio, $N_{\sigma}/N_{\tau}=4.$ Gauge configurations were generated using a modified version of the \texttt{SIMULATeQCD} software suite~\cite{HotQCD:2023ghu} and stored every tenth simulation time unit. We consider physical strange quark mass, $m_s$, and degenerate light quark masses, $m_u=m_d=m_s/27$, which at vanishing magnetic fields yields physical pion mass, $m_{\pi} \simeq 135 ~{\rm MeV}$. The magnetic field strength $eB$ ranges as strong as $\simeq 45M_\pi^2$, corresponding to the flux $N_b$ varying from 1 to 32, such that discretization error in $eB$ is mild, $N_b/N_\sigma^2\ll 1$ \cite{Endrodi:2019zrl}. For the case of $N_b=0$, we adopted lattice QCD results obtained in~\cite{Bollweg:2021vqf}. Our temperature range is focused around the $T_{pc}$ in the considered $eB$ range, $T\in \left[145 - 165\right] {\rm MeV} $. For lattice data at $N_{\tau} = 8$ and $N_{\tau} = 12$, we perform two-dimensional B-spline fits following an approach similar to Ref. \cite{Bali:2011qj} and compute continuum estimates via a joint fit using quadratic extrapolation in $1/N_{\tau}$.

\section{Strangeness neutrality and isospin asymmetry}
\label{sec:initial_cond}

Even the leading-order Taylor expansion of the EoS has several free control parameters, mainly the physical conserved charge potentials. These chemical potentials are interrelated and hence one can express the electric charge and strangeness potentials as a series expansion in terms of the baryon counterpart,
\begin{align}
\label{eqn:muQmuS}
\hat{\mu}_{\tQ}(T,eB,\hat{\mu}_{\tB}) = q_1 (T,eB) \hat{\mu}_{\tB} + \mathcal{O}(\hat{\mu}_{\tB}^3),\quad
\hat{\mu}_{\tS}(T,eB,\hat{\mu}_{\tB}) = s_1 (T,eB) \hat{\mu}_{\tB} + \mathcal{O}(\hat{\mu}_{\tB}^3),
\end{align}
where $q_1 (T,eB), s_1 (T,eB)$ imply the leading-order ratio of  $\hat{\mu}_{\tQ}/\hat{\mu}_{\tB}$ and $\hat{\mu}_{\tS}/\hat{\mu}_{\tB}$ respectively. Additionally, in heavy-ion collision experiments, the initial nuclei colliding conditions impose strangeness neutrality and isospin asymmetry constraints on conserved charges \cite{Bazavov:2012vg, Bazavov:2014xya,Fukushima:2016vix}, and can be used to quantify $q_1, s_1$,
\begin{align}
\langle n_{\tS}\rangle = 0 ~\xrightarrow{}& ~s_1 = - {\left( \chi_{11}^{\tB \tS} + q_1 \chi_{11}^{\tQ \tS}\right)}/ { \chi_{2}^{\tS}}\,, \\
\quad \langle n_{\tQ}\rangle / \langle n_{\tB}\rangle = r ~\xrightarrow{} &~ q_1 = \frac{r \left( \chi_{2}^{\tB}  \chi_{2}^{\tS} - \chi_{11}^{\tB \tS} \chi_{11}^{\tB \tS} \right) - \left(\chi_{11}^{\tB \tQ} \chi_{2}^{\tS} - \chi_{11}^{\tB \tS} \chi_{11}^{\tQ \tS} \right)}{\left(\chi_{2}^{\tQ } \chi_{2}^{\tS} - \chi_{11}^{\tQ \tS}  \chi_{11}^{\tQ \tS}\right) - r\left( \chi_{11}^{\tB \tQ} \chi_{2}^{\tS} -\chi_{11}^{\tB \tS} \chi_{11}^{\tQ \tS} \right)}\,,
\end{align}
where $r$ is the isospin parameter. For $Pb$/$Au$ heavy-ion collisions with quark content $u:d:s \equiv 46.5:53.5:0$, we expect $r =0.4$, implying slight isospin asymmetry.

\begin{figure}[htbp]
\centering

\includegraphics[width=0.35\textwidth]{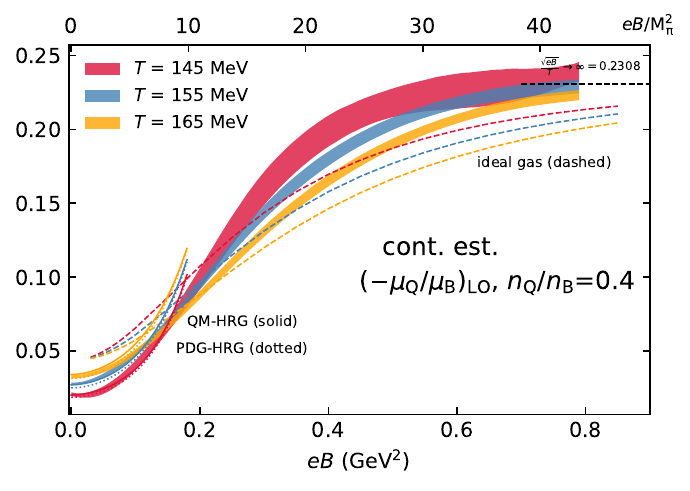} \quad
\includegraphics[width=0.35\textwidth]{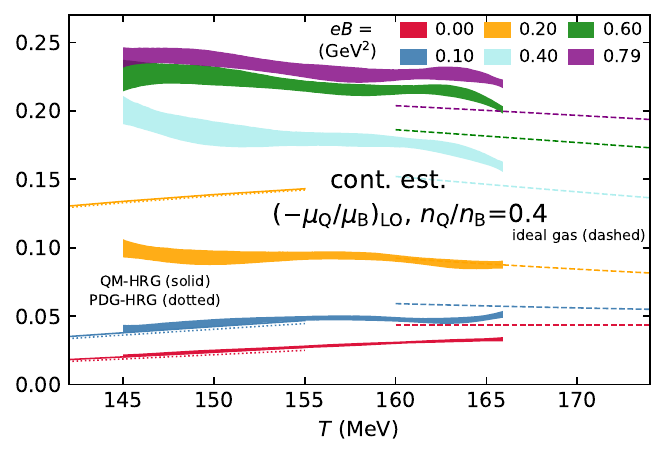}

\includegraphics[width=0.35\textwidth]{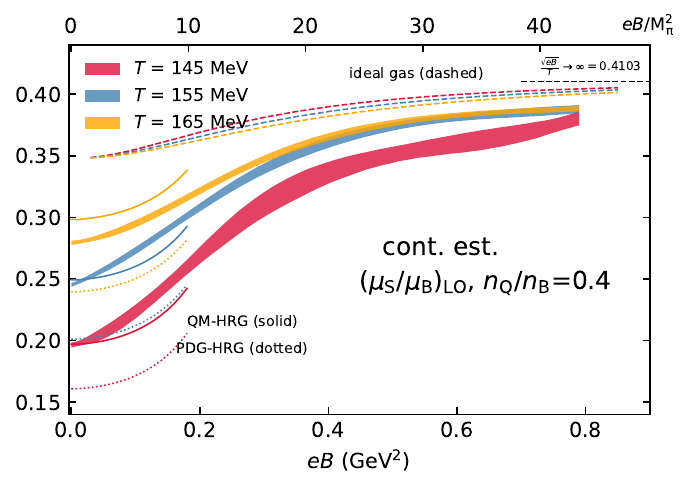} \quad
\includegraphics[width=0.35\textwidth]{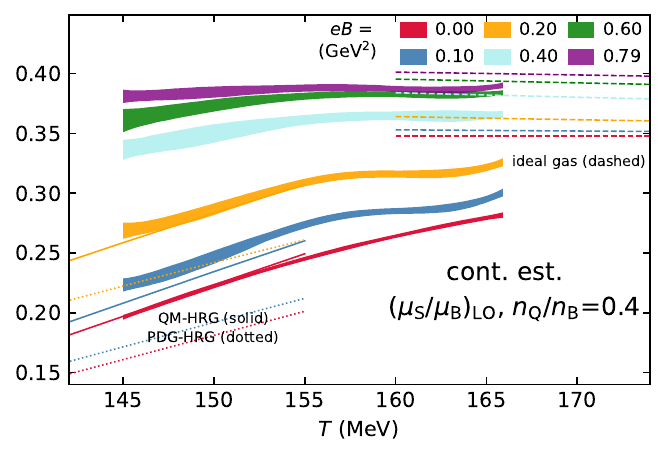}

\caption{Leading-order coefficients $q_1$ (top) and $s_1$ (bottom) as functions of $eB$ (left) and $T$ (right).}
\label{fig:q1s1}
\end{figure}

In \autoref{fig:q1s1}, we show the leading-order coefficients $q_1$ and $s_1$ as functions of $eB$ at fixed $T$ values and vice versa, highlighting their behaviour around the $T_{pc}$ region. Let us try to understand the behaviour of $q_1$. Apparently, $q_1$ is negative throughout the $T-eB$ range even for $eB=0$. In our choice of collision system with slight isospin asymmetry, $r=0.4$ implies an unequal number of neutral and charged particles, which demands a suppression in the content of positively charged protons, which in fugacity terms implies charged contribution be greater than neutral counterpart $e^{-Q_i \hmuQ}> e^0$. This consequently constrains $\hmuQ < 0$, and hence $q_1$ is negative. Interestingly, upon the introduction of magnetic fields, $q_1$ becomes more negative. In the context of the HRG model, the thermal (neglecting vacuum energy terms) pressure arising from charged particles of mass $m_i$, spin $s_z$ in a magnetic field can be expressed as
\begin{align}
\label{eqn:hrg}
\frac{p_c^{M/B}}{T^4} =\frac{|q_i|B}{2\pi^2 T^3} \sum_{s_z =-s_i}^{s_i}   \sum_{l=0}^{\infty}     \epsilon_0  \sum_{k=1}^{\infty} \left(\pm 1\right)^{k+1} \frac{e^{k{\mu_i}/T}}{k}  {\rm K}_1 \left( \frac{k \epsilon_0}{T} \right) \\
   {\rm with}~\epsilon_0  = \sqrt{m_i^2 + 2 |q_i| B (l+1/2-s_z)} \,,
\end{align}
where the perpendicular motion is quantized into Landau levels $l$ (see Ref. \cite{Fukushima:2016vix,Ding:2021cwv,Ding:2023bft}). Here, $\text{K}_1$ and $\text{K}_2$ are the first- and second-order modified Bessel functions of the second kind, respectively. Both PDG-HRG and QM-HRG are in reasonable agreement with lattice continuum estimates, suggesting that, to some extent, the HRG model provides a reasonable physical interpretation for relatively weaker-$eB$ and low-$T$. Apparently for $eB\neq0$, the lowest Landau level occupation increases with decreasing energy, resulting in positively charged protons being more favoured, however, the fixed isospin parameter constraint becomes even stricter, enforcing more negative $\mu_{\rm Q}$ and $q_1$.  \footnote{Neutral particles are considered immune to such effects. Although, in principle, we know both protons and neutrons have non-trivial (nonzero) magnetic moments which is beyond HRG interpretation.} Around $eB \sim 0.2~{\rm GeV}^2$ HRG begins to overshoot Lattice results and eventually breaks down. Interestingly, as $eB$ grows further $eB\gtrsim 0.2~{\rm GeV}^2$, the dependence $q_1$ on $T$ and $eB$ changes drastically. We observe crossings between the fixed $T$ bands in the $eB$ dependence which are attributed to non-monotonic behaviour in the $T$ dependence, implying a non-trivial change in the degrees of freedom in the presence of strong magnetic fields. In the extremely strong-$eB$ sector with $eB\sim 0.8 ~{\rm GeV}^2$, we observe saturation to magnetized ideal gas expressed as 
\begin{align}
    \label{eqn:ideal}
\hat{p} \equiv \frac{p}{T^4} &= \frac{8\pi^2}{45} + \sum_{f=u,d,s} \frac{3 \left| q_f\right| B}{\pi^2 T^2} \left[ \frac{\pi^2}{12} + \frac{\hat{\mu}^2_f}{4} +p_f(eB) \right] \\
   {\rm with}~ p_f(eB) &= 2\frac{\sqrt{2 \left| q_f \right| B  }}{T} \sum_{l=1}^{\infty} \sqrt{l} \sum_{k=1}^{\infty} \frac{(-1)^{k+1}}{k} ~\cosh(k \hat{\mu}_f)
 {\rm K}_1\left( \frac{k\sqrt{2\left| q_f \right| Bl }}{T} \right)\,,
\end{align}
where ${\hat{\mu}^2_f}$ corresponds to LLL contribution and $p_f(eB)$ to higher levels (more details in Ref. \cite{Ding:2021cwv}). Note that, $q_1$ is a ratio observable, and from the ideal gas approximation, the above linear dependence on the magnetic field cancels out. This results in a clear saturation effect corresponding to the free limit, where  $\sqrt{eB}/T \to \infty$. Similar interpretations and arguments for enhancements apply to $s_1$ but with an opposite sign since the strange quarks inherently carry negative strangeness.

\section{Leading-order magnetic EoS}
\label{sec:results}

The thermal pressure of QCD magnetic EoS, using \autoref{eqn:muQmuS} from the previous section to help constrain the Taylor expansion, can be expressed as
\begin{align}
\label{}
\hat{p} (T, eB,\hmuB)
=\sum_{ijk}\frac{1}{i!j!k!}~ \chi^{\tB \tQ \tS}_{ijk}~ \hat{\mu}^{i}_{\tB}  \left(  q_1 (T) \hat{\mu}_{\tB} + \mathcal{O}( \hat{\mu}_{\tB}^3) \right)^{j} \left(  s_1 (T) \hat{\mu}_{\tB} + \mathcal{O}( \hat{\mu}_{\tB}^3)\right)^{k},\\
\text{series in $\hmuB$,}\quad \Delta\hat{p}\equiv \hat{p}(T,eB,\mu_{\tB}) -\hat{p}(T,eB,0) 
= \sum_{k=1}^{\infty} P_{2k}(T,eB) \hat{\mu}_{\tB}^{2k}.
\end{align}
where for $k=1$, $P_{2}$ is the leading-order Taylor coefficient for pressure
\begin{align}
P_2 &= \frac{1}{2!}\left( \chi^{\tB}_{2}  + \chi^{\tQ}_{2} q_1^2 +  \chi^{\tS}_{2} s_1^2 \right)  + \chi^{\tB \tQ}_{11} q_1 + \chi^{\tB \tS}_{11}  s_1+ \chi^{\tQ \tS}_{11} q_1 s_1
\end{align}
with initial nuclei conditions hidden in leading-order ratios $q_1$ and $s_1$. Changes in pressure coefficients inherently characterize the changes in dominant degrees of freedom in the thermodynamical system. At vanishing magnetic fields, it is well established that the pressure increases monotonically as the temperature rises due to the thermal agitations, with a pronounced rise in the QCD transition region. In the presence of magnetic fields, the behaviour of pressure is much more intricate due to the interplay between thermal and magnetic field effects.

\begin{figure}[htbp]
\centering

\includegraphics[width=0.37\textwidth]{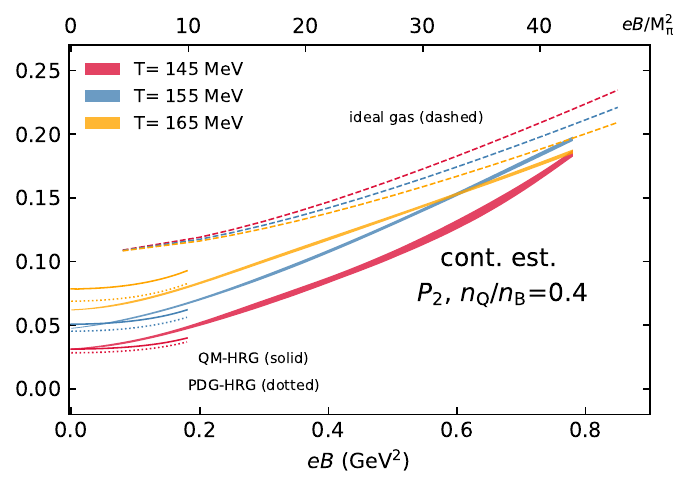}
\includegraphics[width=0.36\textwidth]{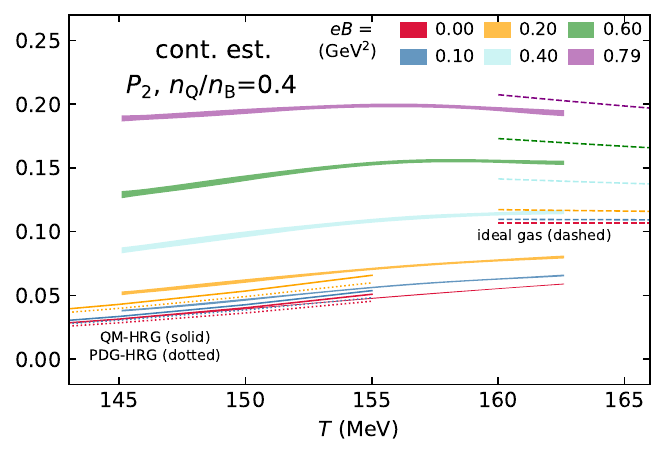}

\includegraphics[width=0.24\textwidth]{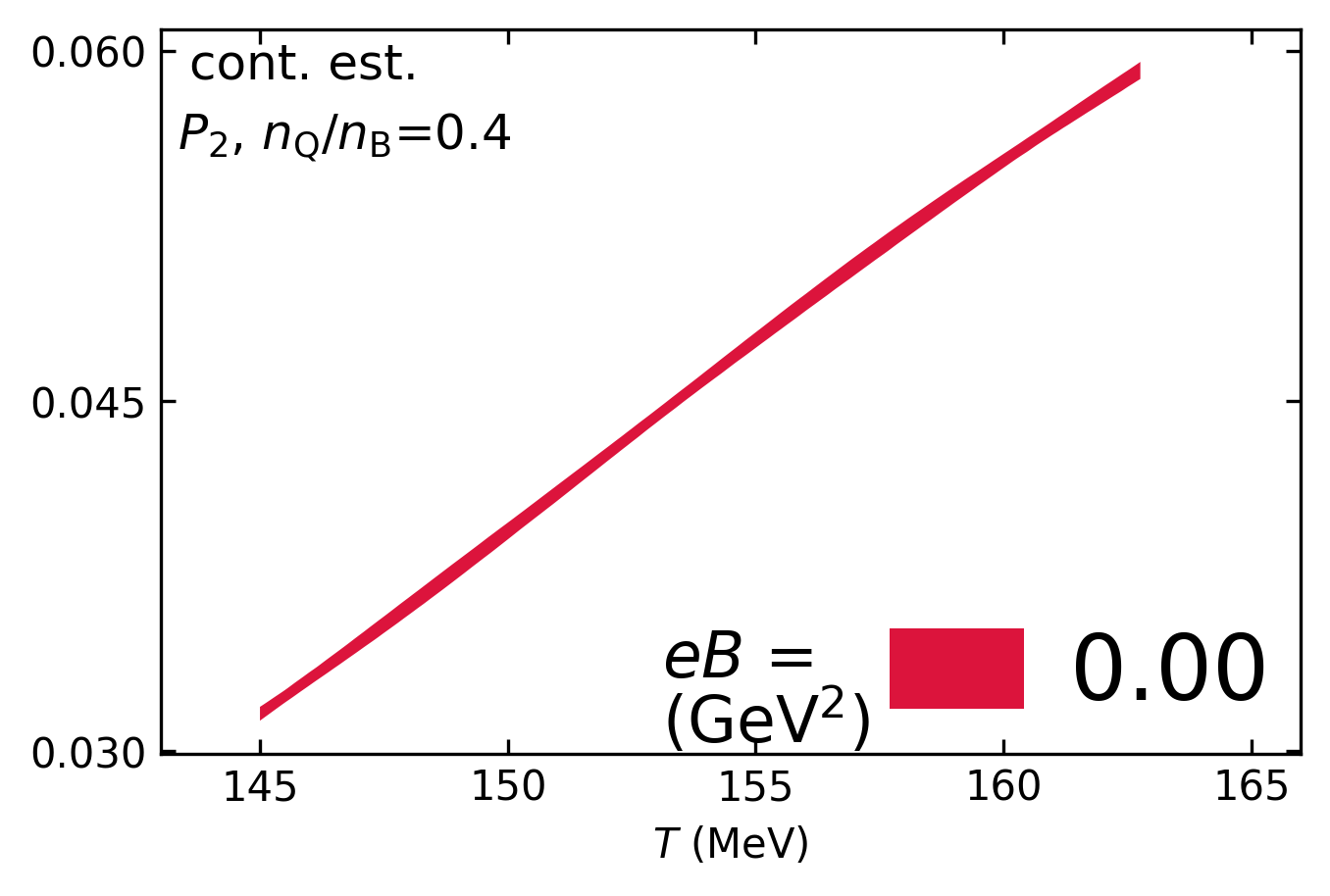}
\includegraphics[width=0.24\textwidth]{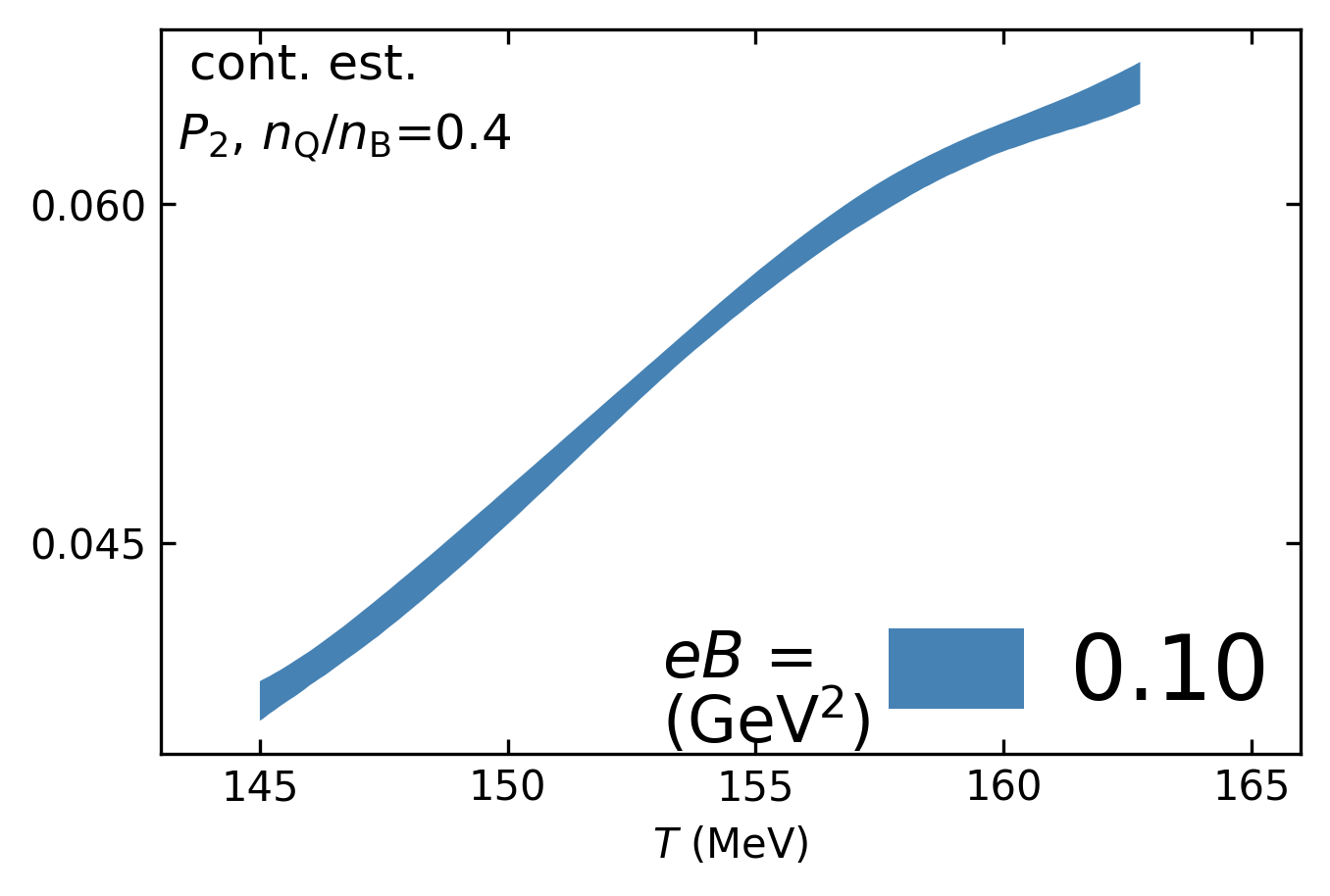}
\includegraphics[width=0.24\textwidth]{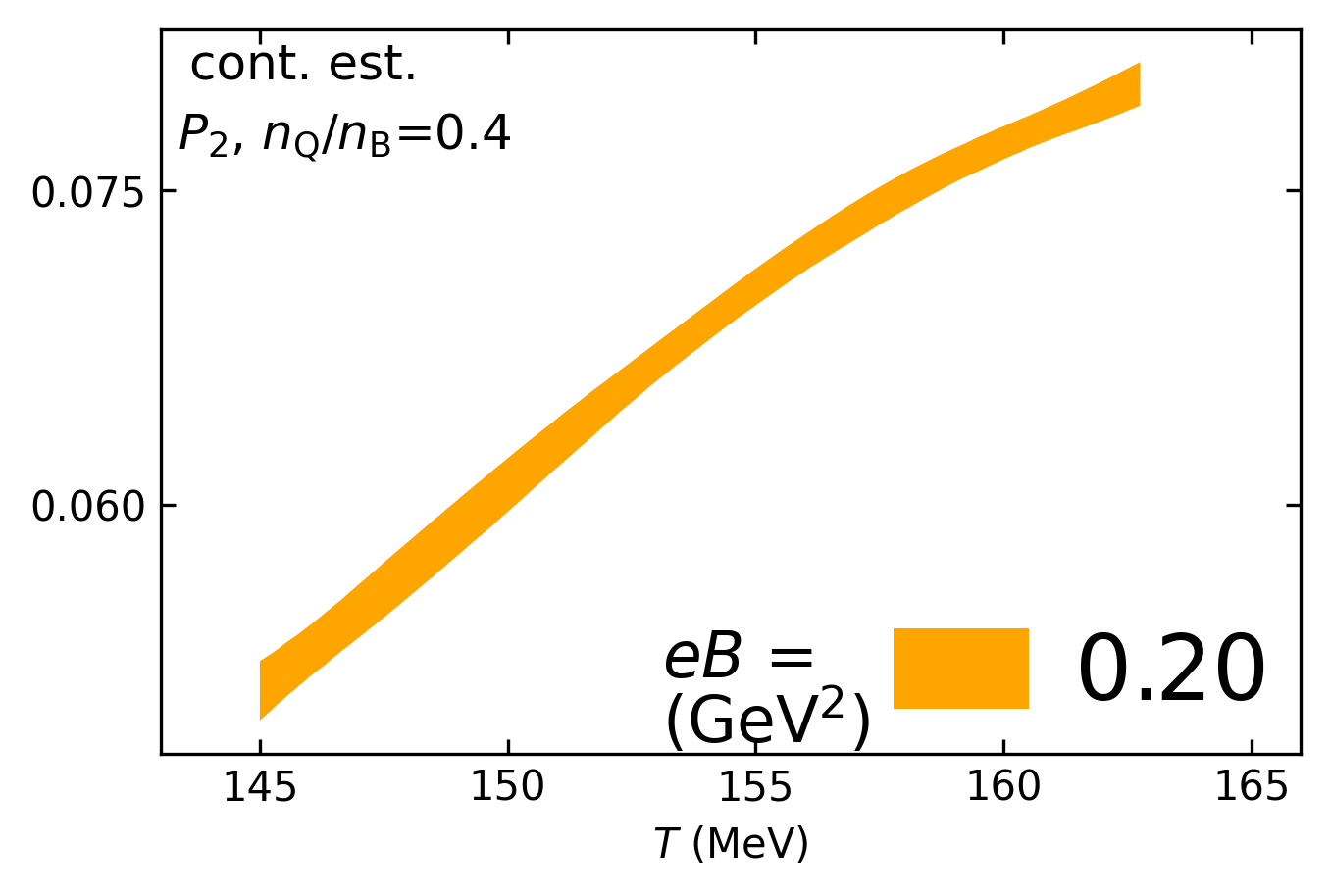}

\includegraphics[width=0.24\textwidth]{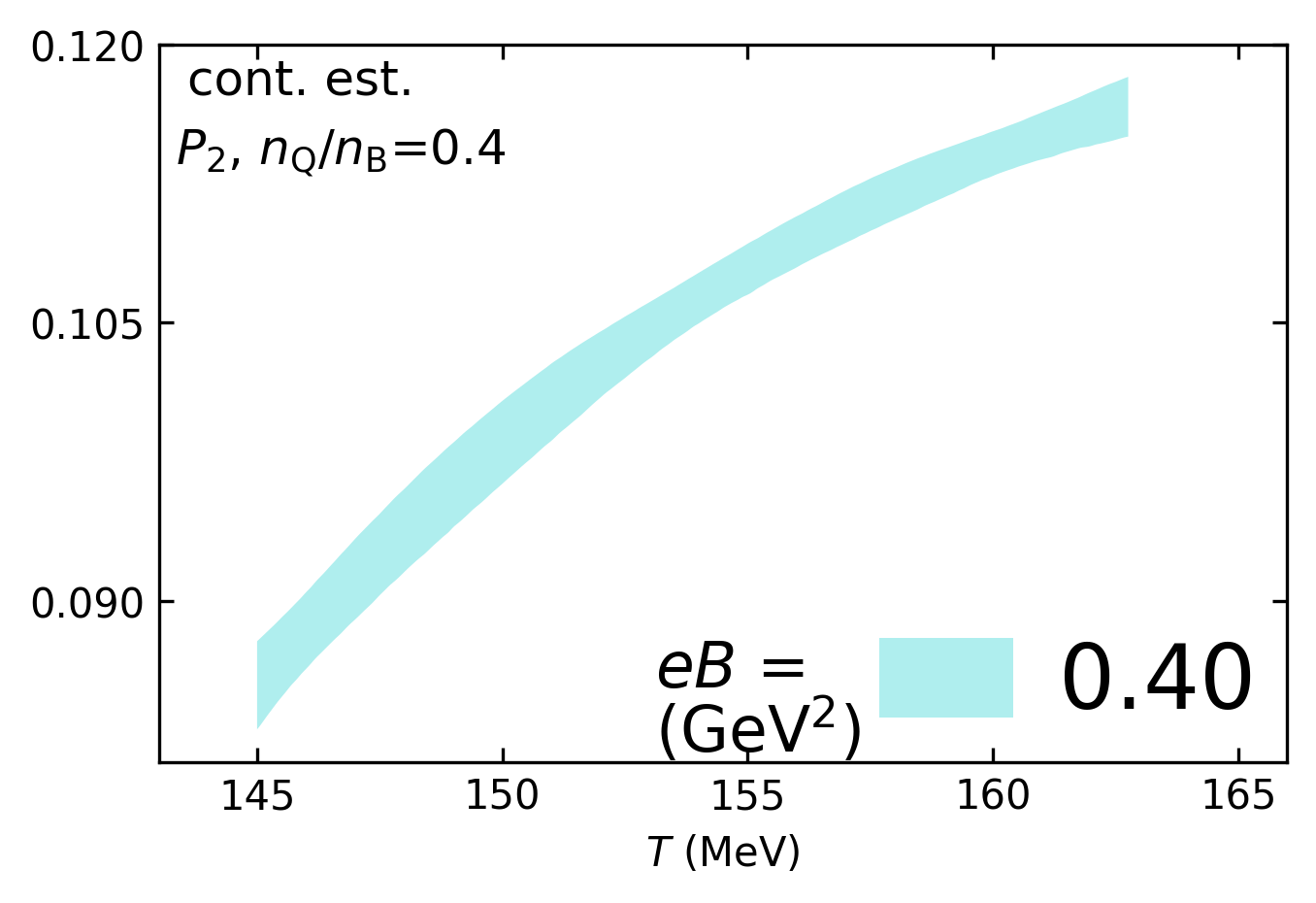}
\includegraphics[width=0.24\textwidth]{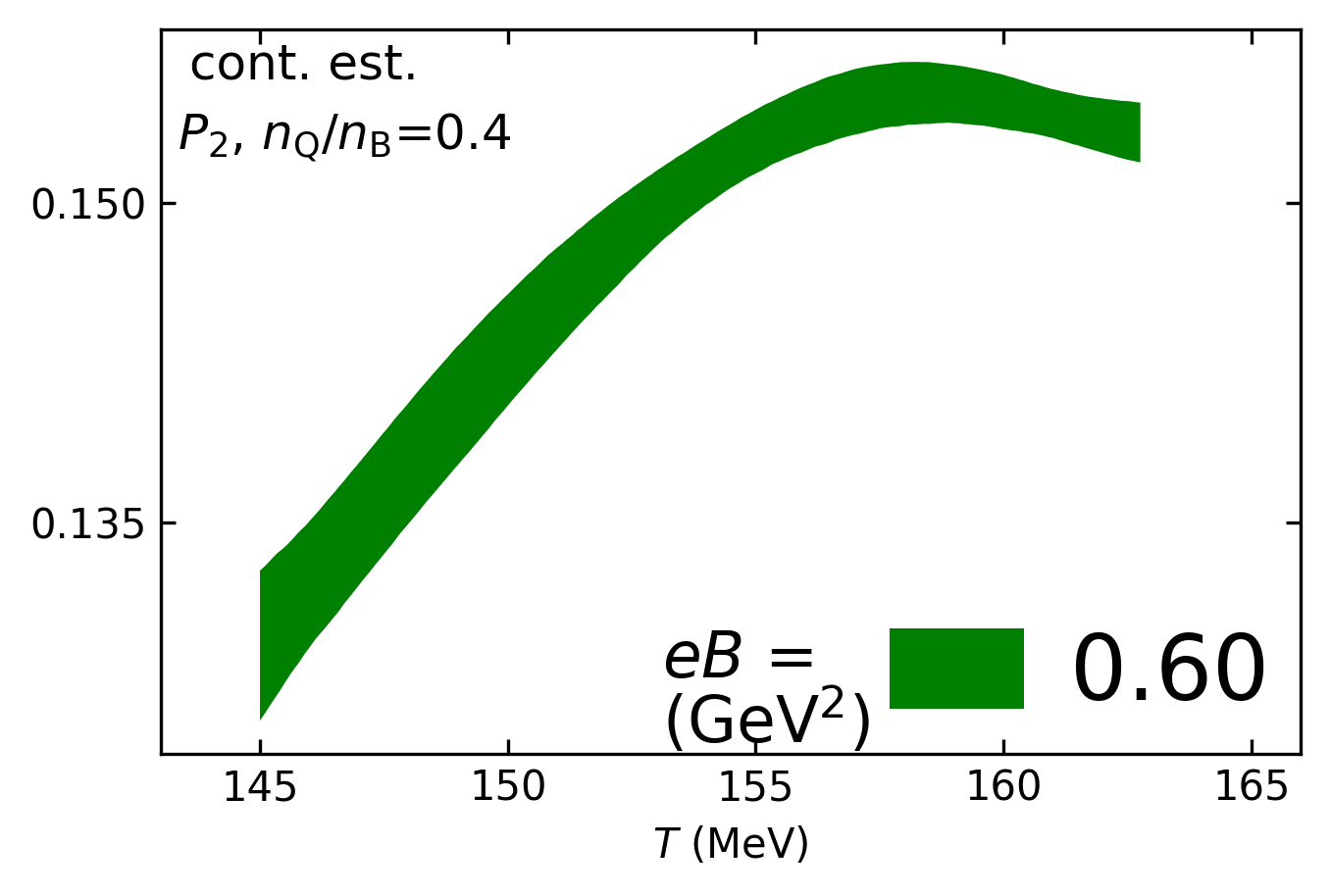}
\includegraphics[width=0.24\textwidth]{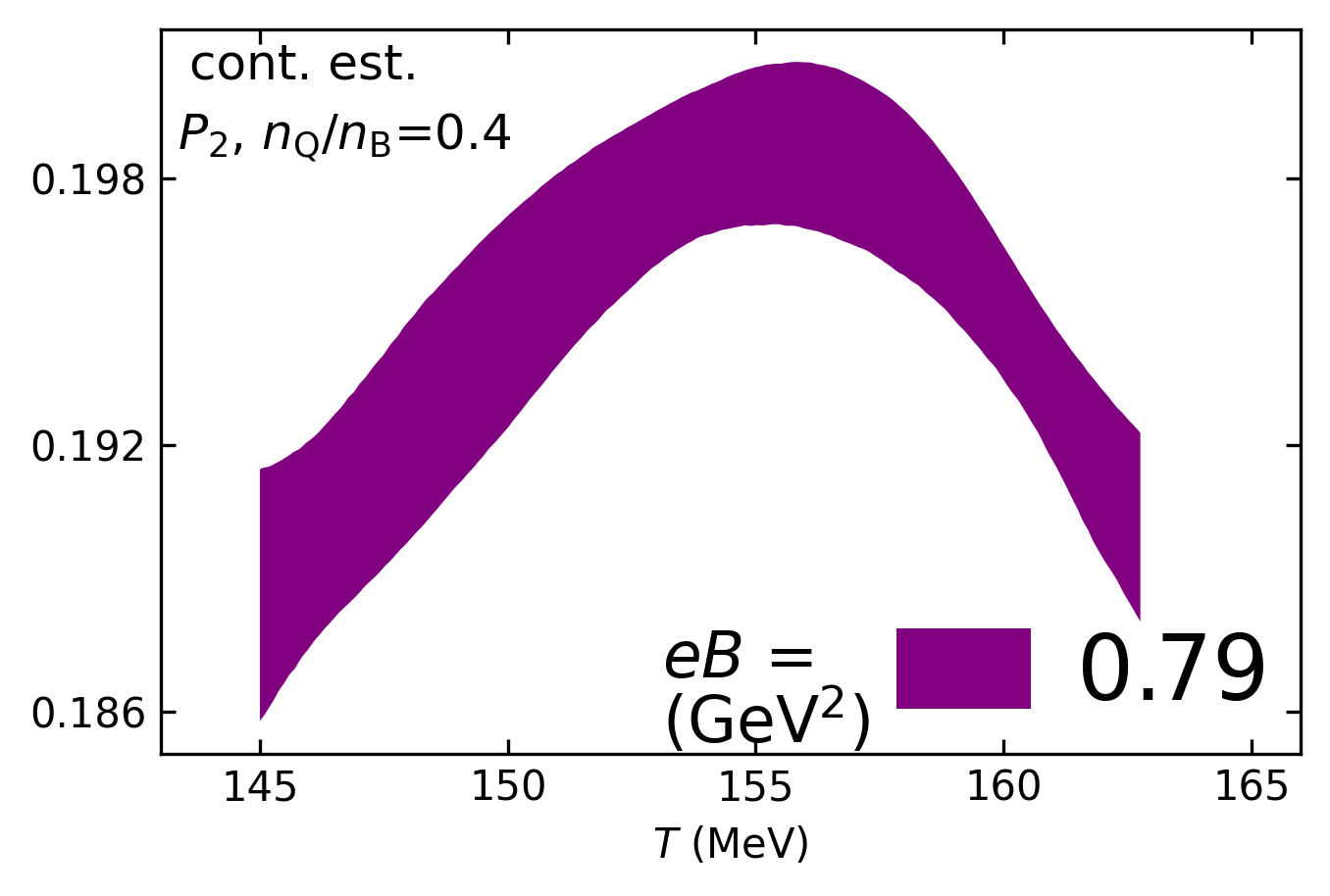}

\caption{Leading-order coefficients of pressure difference $P_2$ as function of $eB$ and $T$.}  
\label{fig:vseB/p2_Pb_Au_cont}
\end{figure}

In \autoref{fig:vseB/p2_Pb_Au_cont}, on the top-left and top-right panels, we plot continuum estimates of $P_2$ as a function of $eB$ and $T$, respectively, around $T_{pc}(eB=0)$ region. As seen in the previous section, HRG interpretation of lattice data is only applicable for the relatively weaker-$eB$ and low-$T$ regime. As the strength $eB$ grows, the pressure keeps increasing, stemming primarily from the fact that the degeneracy of Landau levels is directly proportional to the field strength. Notice that the pressure coefficients are connected to the fluctuations of conserved charges and recent lattice works have studied in detail how magnetic field enhances these fluctuations \cite{Ding:2023bft,Ding:2021cwv,Borsanyi:2023yap,Astrakhantsev:2024mat}.  In an extremely strong-$eB$ regime, the lattice data begins to approach magnetized ideal gas. Unlike the ratio observables $q_1$ and $s_1$, there is no saturation approach with increasing $eB$, for pressure. It can be understood from Ref. \cite{Ding:2021cwv}, that free limit saturation for such strong magnetic fields is expected at very high temperatures.
What is further interesting is the signs of crossing among fixed temperature bands in the strong-$eB$ regime, these crossings imply that the functional dependence is significantly altered by strong magnetic fields. The plots in the bottom panel highlight the drastic change in the temperature dependence of pressure for different fixed strength $eB$ values, with non-monotonic behaviour in $T$-dependence for $eB>0.6~ {\rm GeV}^2$, and formation of peak structures around $eB=0.79~ {\rm GeV}^2$.

\begin{figure}[htbp]
\centering

\includegraphics[width=0.34\textwidth]{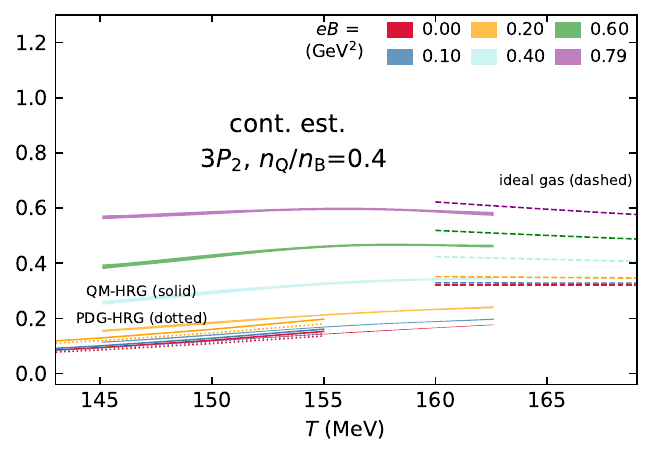}
\includegraphics[width=0.32\textwidth]{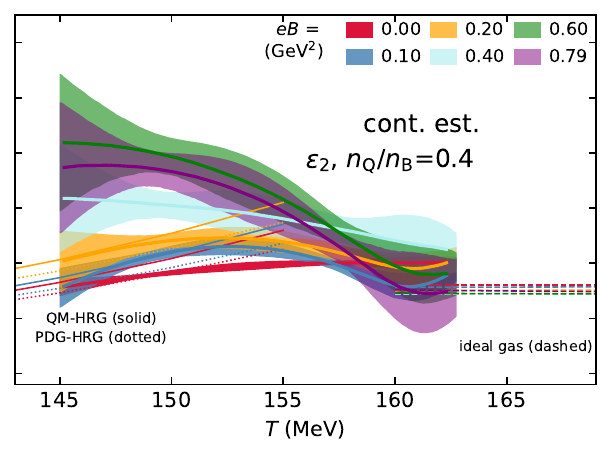}
\includegraphics[width=0.32\textwidth]{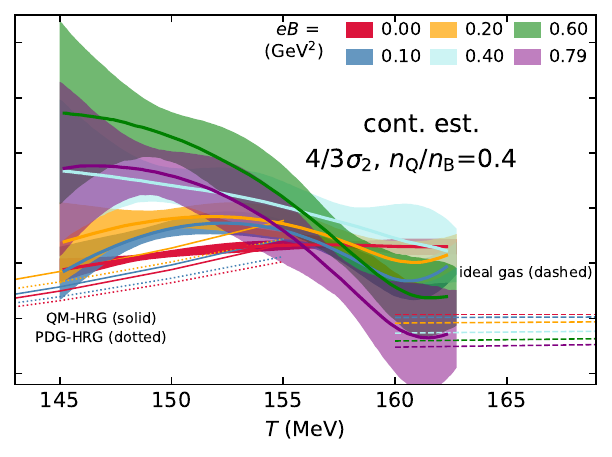}
\caption{Leading-order coefficients of pressure 
$P_2$ (left), energy density $\epsilon_2$ (middle), and entropy density $\sigma_2$ (right) as function of temperature.}  
\label{fig:vsT/eps2_sig2_Pb_Au_cont_vsT}
\end{figure}

To define higher-order bulk observables like energy $\epsilon$ and entropy $\sigma$ densities, we need to compute temperature derivatives of the expansion coefficients entering the Taylor series of $\hat{p}$, $\Xi^{\rm BQS}_{ijk}(T) = T~\partial \chi^{\tB \tQ \tS}_{ijk}/ \partial T$ such that,
\begin{align}
        \label{eqn:eps2}
    {\hat{\epsilon}} (T,eB, \mu) \equiv {\hat{\Theta}+3\hat{p}} = \sum_{ijk}\frac{1}{i!j!k!}~\left(\Xi^{\tB \tQ \tS}_{ijk} + 3\chi^{\tB \tQ \tS}_{ijk} \right) \hat{\mu}^{i}_{\tB} \hat{\mu}^{j}_{\tQ} \hat{\mu}^{k}_{\tS},\\
    \label{eqn:sig2}
    {\hat{\sigma}} (T,eB, \mu)\equiv {\hat{\epsilon}+\hat{p} - \sum_{\mathcal{C}} \hat{\mu}_\mathcal{C} \hat{n}^{\mathcal{C}}}  = \sum_{ijk}\frac{1}{i!j!k!}~\big(\Xi^{\tB \tQ \tS}_{ijk} + \left[ 4-(i+j+k)\right] \chi^{\tB \tQ \tS}_{ijk} \big)
     \hat{\mu}^{i}_{\tB} \hat{\mu}^{j}_{\tQ} \hat{\mu}^{k}_{\tS}.   
\end{align}
The above relations incorporating trace anomaly, $\hat{\Theta} (T,eB, \mu)$ provide insights into the non-vanishing trace of the energy-momentum tensor arising from quantum interactions and scale dependent renormalization. 
$\hat{n}^{\mathcal{C}} \equiv {n^\mathcal{C}}/{T^3}={\partial \hat{p}}/{\partial \hat{\mu}_{\mathcal{C}}}$ denotes conserved charge densities. Following the similar routine as discussed for pressure, we express in terms of $\hmuB$ as 
\begin{align}
\Delta \hat{\epsilon}  = \sum_{m=1}^{\infty} \epsilon_{2m} \hmuB^{2m},\quad
\Delta \hat{\sigma}= \sum_{m=1}^{\infty}\sigma_{2m} \hmuB^{2m}, \quad 
\hat{n}^{\rm B} = \sum_{k=1}^{\infty} N^{\rm B}_{2k-1} \hat{\mu}_{\tB}^{2k-1}\end{align}
and setting $k,m=1$ for the leading-order we arrive at
\begin{align}
\epsilon_{2} \equiv \epsilon_{2} (T,eB) = 3P_2 -rT~\partial_T q_1 N^{\rm B}_1   +~T ~\partial_T P_2\,,\\
\sigma_{2} \equiv \sigma_{2}(T,eB) = \epsilon_{2} +P_2 - (1+rq_1)N^B_1\,.
\end{align}
The expansion coefficients for energy and entropy densities exhibit a more intricate dependence on temperature and magnetic field than those for pressure. These leading-order coefficients, involving temperature derivatives, provide crucial insights into the slopes of $P_2$ and $q_1$ and higher-order derivatives such as $P_4$.  In \autoref{fig:vsT/eps2_sig2_Pb_Au_cont_vsT}, we show together the leading-order pressure ($P_2$), energy ($\epsilon_2$) and entropy ($\sigma_2$) densities. Despite significant errors, the figure reveals that as $eB$ increases from 0 to 0.6 GeV$^2$, both densities rise, peaking at $eB \sim 0.6$ GeV$^2$, beyond which they decline. This peak coincides with the onset of non-monotonicity in $P_2$, qualitatively suggesting a link between the peak in $P_2$ and the reduction in $\epsilon_2$ and $\sigma_2$.

\section{Summary \label{sec:conclusion}}

In this work, we explored the (2+1)-flavor QCD equation of state in the presence of strong magnetic fields from first-principles lattice computations. Our simulations employed highly improved staggered quarks at the physical pion mass on $32^3 \times 8$ and $48^3 \times 12$ lattices and continuum estimates were taken. We considered magnetic field strengths ranging up to 0.8 GeV$^2$ with the temperature window focused around $T_{pc}$. We first determined the leading-order coefficients $q_1$ and $s_1$ which directly encode the strangeness neutral and isospin asymmetry conditions relevant for heavy-ion collisions. Notably, we observed crossings in the fixed-$T$ bands in $eB$-dependence and saturation of $q_1$ and $s_1$ in extremely strong-$eB$ regimes. We also sketched the physical descriptions of such behaviour and found that the HRG model breaks down in a strong-$eB$ regime, while the magnetized ideal gas can serve as a high-$T$, strong-$eB$ reference.

Later we utilized these constraints to analyze the leading-order behaviour of pressure Taylor coefficients in the presence of magnetic fields. Interestingly, under these strong magnetic fields, we observed mild peak formation implying a non-trivial change in degrees of freedom, along with drastic modification in the $T$-dependence of pressure and hints towards $T_{pc}$ lowering. Furthermore, we explored the thermodynamics of energy and entropy density in magnetic fields, observing an initial increase with $eB$ followed by a peak around $eB \sim 0.6~{\rm GeV}^2$ and a subsequent decline. These observations highlight the need for further explorations in the presence of magnetic fields to unveil the non-perturbative origin of such enhancements and non-monotonic behaviours.

\acknowledgments

This work is supported partly by the National Natural Science Foundation of China under Grants No. 12293064, No. 12293060, and No. 12325508, as well as the National Key Research and Development Program of China under Contract No. 2022YFA1604900. The numerical simulations have been performed on the GPU cluster in the Nuclear Science Computing Center at Central China Normal University ($\mathrm{NSC}^{3}$) and Wuhan Supercomputing Center.

\bibliographystyle{JHEP.bst}
\bibliography{ref.bib}

\end{document}